\documentclass[aps,prb,twocolumn,superscriptaddress,showpacs]{revtex4}

\usepackage{amsmath,epsfig,color,amssymb}
\usepackage{graphicx}
\usepackage{graphicx}
\usepackage{dcolumn}
\usepackage{bm}
\usepackage{graphicx}
\usepackage{amsmath, amsfonts, amssymb,mathrsfs}
\usepackage{pstricks}
\usepackage{amsxtra}
\usepackage{amsthm}
\usepackage{natbib}

\makeatletter

\def\be{\begin{equation}}
\def\ee{\end{equation}}
\def\bea{\begin{eqnarray}}
\def\eea{\end{eqnarray}}
\newcommand{\ket}[1]{\mbox{$|#1\rangle$}}
\newcommand{\bra}[1]{\mbox{$\langle#1|$}}

\def\br{{\bm{r}}}

\def\be{\begin{equation}}      
\def\ee{\end{equation}}
\def\beu{\begin{equation*}}   
\def\eeu{\end{equation*}}

\providecommand{\abs}[1]{\left\lvert#1\right\rvert}   
\providecommand{\ket}[1]{\left|#1\right\rangle}

\providecommand{\bra}[1]{\left\langle#1\right|}

\providecommand{\del}{\partial}

\graphicspath{../}

\begin{document}

\title{Optical Control of Donor Spin Qubits in Silicon}
\author{M.~J.~Gullans}
\affiliation{Joint Quantum Institute, National Institute of Standards and Technology, Gaithersburg, MD 20899, USA}
\affiliation{Joint Center for Quantum Information and Computer Science, University of Maryland, College Park, MD 20742, USA}
\author{J.~M.~Taylor}
\affiliation{Joint Quantum Institute, National Institute of Standards and Technology, Gaithersburg, MD 20899, USA}
\affiliation{Joint Center for Quantum Information and Computer Science, University of Maryland, College Park, MD 20742, USA}
\date{\today}


\begin{abstract}
We show how to achieve optical, spin-selective transitions from the ground state to excited orbital states of group-V donors (P, As, Sb, Bi) in silicon.
We consider two approaches based on either resonant, far-infrared (IR) transitions of the neutral donor or resonant, near-IR excitonic transitions.  For far-IR light, we calculate the dipole matrix elements between the  valley-orbit and spin-orbit split states for all the goup-V donors using effective mass theory.  We then calculate the maximum rate and amount of electron-nuclear spin-polarization achievable through optical pumping with circularly polarized light.  We find this approach is most promising for Bi donors due to their large spin-orbit and valley-orbit interactions.
Using near-IR light, spin-selective excitation is possible for all the donors by driving a two-photon $\Lambda$-transition from the ground state to higher orbitals with even parity.  We show that externally applied electric fields or strain allow similar, spin-selective $\Lambda$-transition to odd-parity excited states.    
We anticipate these results will be useful for future spectroscopic investigations of donors, quantum control and state preparation of donor spin qubits, and for developing a coherent interface between donor spin qubits and single photons.  
\end{abstract}
\maketitle

\section{Introduction}
The optical spectroscopy and control of shallow group-V donors (P, As, Sb, Bi) in silicon has a long history. \cite{Ramdas81,Davies89}
Recent work has focused on achieving optical control of the group-V donors for applications to quantum information\cite{Zwanenburg13} and far-infrared (IR) lasers.\cite{Pavlov12}  One common approach takes advantage of the excited orbital states of the neutral donors (D$^0$ states), which have hydrogen-like $s$-, $p$-, $d$-,~etc.\ orbitals.\cite{Kohn55}
Resonant excitation into these states allows one to turn on and off interactions between donors by exciting and de-exciting the system\cite{Stoneham03,Vinh08,Greenland10} or to create population inversion for lasing.\cite{Pavlov12,Zhukavin11}  

The second class of approaches makes use of the neutral, donor bound exciton state consisting of a hole bound to the doubly occupied donor (D$^0$X states).\cite{Davies89}  This state is appealing to use for optical control because it has a long lifetime, a large spin-orbit splitting, and an easily accessible transition frequency in the near-infrared.\cite{Yang06}   It has been successfully employed for electron and nuclear spin state preparation and measurement and spin-to-charge conversion for spin readout of P and Bi donor ensembles.\cite{Yang06,Yang09,Sekiguchi10,Saeedi15,Lo15}  


\begin{figure}[thb]
\begin{center}
\includegraphics[width=.43 \textwidth]{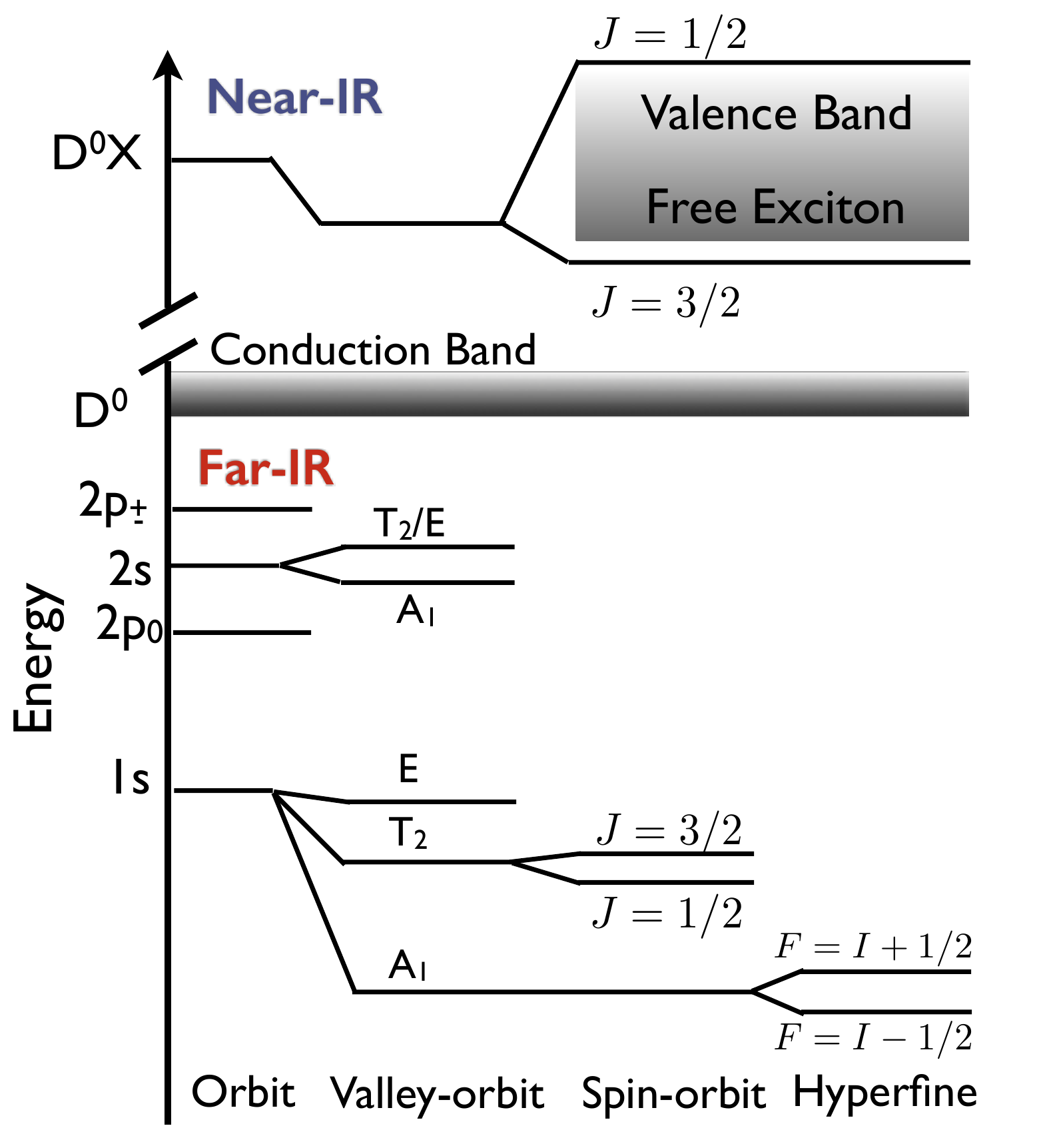}
\caption{Energy scales in optical spectroscopy of group-V donors in silicon.  Here $A_1$, $T_2$ and $E$ refer to irreducible representations of the tetrahedral group, $J$ is the total angular momentum quantum number,   $F$ is the hyperfine quantum number, and $I$ is the total nuclear spin quantum number.}
\label{fig:1}
\end{center}
\end{figure}

In this Article, we describe several approaches to achieve optical, spin-selective excitation from the ground state to the excited orbital states.  Figure \ref{fig:1} illustrates the energy scales and terms relevant for optical spectroscopy and control of the group-V donors.   
We first analyze control schemes based on near resonant, far-IR transitions.  These transitions are most promising for Bi donors because of their large spin-orbit  and valley-orbit interactions.  Using multi-valley effective mass theory,\cite{Ning71,Shindo76,Tyuterev10,Pica14a,Mortemousque15,Smit04,Friesen05,Pica14b,Gamble14} we calculate the dipole matrix elements between the  spin-valley-orbital states for all the donors and estimate the maximum rate and amount of electron-nuclear spin-polarization achievable with this technique.
We then consider optical approaches in the near-IR using the D$^0$X state.  In this case, spin-selective excitation is possible for all the donors by exploiting a two-photon  $\Lambda$-transition from the ground state to higher orbitals with even parity (e.g., $2s$, $3d$).  Using externally applied electric fields or strain, we show it is also possible to achieve similar, spin-selective  $\Lambda$-transitions to odd-parity (e.g., $2p$) excited states.  


 The paper is organized as follows.  In Sec.\ \ref{sec:2} we review the effective mass theory (EMT) used to describe the donor orbital wavefunctions and the donor bound exciton state.  
In Sec.\ \ref{sec:3} we use EMT 
to calculate parity forbidden dipole matrix elements between spin-valley-orbital states of the donors.
To take into account the large corrections from the donor impurity potential for the $1s$ states, we use the recently developed variational solutions provided in Ref. \onlinecite{Pica14a}.
Using these calculated dipole elements, we consider the achievable amount and rate of electron-nuclear spin polarization via optical pumping through the spin-orbit split $1sT_2$ states.
In Sec.\ \ref{sec:4} we consider several different schemes to achieve spin-selective excitation to the excited orbital states of the donor through the D$^0$X state.  
We calculate the relative strength of the dipole transition from the D$^0$X state to the even-parity states by using a Hartree-Fock variational solution to the D$^0$X ground state.\cite{Leuenberger09}  We then show how to excite the odd-parity, $2p$ states through two-photon transitions using a static electric field or externally applied strain.  Throughout this work we use simple symmetry arguments and variational solutions to obtain order of magnitude estimates for these effects.  A detailed and accurate understanding of these effects will require a combination of more refined measurements and ab-initio theory.


\section{Effective Mass Theory}
\label{sec:2}
In this section we outline the formulation of the effective mass theory and variational solutions to the donor wavefunction orbitals. 
The donor atom is a point-like defect in the lattice, while the bound electron wavefunction extends across hundreds of lattice sites.  As a result, the combined system probes the silicon lattice at both long and short wavevectors.
Early on it was realized that the excited orbital states cam be understood within a hydrogen-like model and have a universal spectrum.\cite{Kohn55}  The electron in the ground state, however, has a large overlap with the donor atom, resulting in a large, donor-specific valley splitting, i.e., chemical shift. \cite{Kohn55}     The interaction with this core potential results in sufficient complexity that a first-principles understanding of these effects is only now being developed through a combination of scanning tunneling microscopy measurements, atomistic simulations, and variational methods.\cite{Wellard05,Salfi14,Klimeck02,Pica14a,Pica14b,Gamble14,Saraiva15}

In what follows, we assume the validity of
electron-like quasiparticle excitations on the semiconductor vacuum
with a local crystal potential and a static, isotropically screened
Coulomb interaction. This approximation has been very successful in
semiconductors and results in an effective Schr{\"o}dinger equation for the donor electron in the presence of an impurity of the form
\be \label{eqn:H}
 E\, \psi(\br)=\bigg[ -\frac{\hbar^2 \nabla^2}{2 m_0} + V^0(\br) +U(\br)\bigg] \psi(\br),
\ee
where $\hbar$ is Planck's constant, $m_0$ is the electron mass, $V^0(r)$ is the periodic potential of the undoped Si lattice, $U(r)$ is the attractive potential of the impurity, and $E$ is the electron energy.

Si is an indirect band gap semiconductor and, consistent with the tetrahedral symmetry, the conduction band has six degenerate valleys located along the crystallographic $\langle100 \rangle$ directions at the points $\bm{k}_{i0}= k_0 \hat{\imath}$ with $\hat{\imath} =\pm \hat{x},\pm \hat{y}, \pm \hat{z}$ and $k_0 \approx 0.85\times 2\pi/a_{si}$ ($a_{si}=0.543$~nm is the lattice spacing).  

The eigenstates of Eq.\ (\ref{eqn:H}) take the form
\begin{align}
\ket{\psi} &= \sum_\mu \alpha^\mu \int d^3k_\mu F_\mu(\bm{k}_\mu) \ket{\bm{k}_\mu+\bm{k}_{\mu0}}, \\
\langle \bm{r}&\ket{\bm{k}_\mu+\bm{k}_{\mu0}} = u_0(\bm{k}_\mu+\bm{k}_{\mu0},\br)e^{i (\bm{k}_\mu+\bm{k}_{\mu0}) \cdot \br},
\end{align}
where $\mu=\pm x, \pm y, \pm z$ sums over the conduction band minima and $u_0$ is the Bloch wavefunction, which is invariant under lattice translations.  The effective mass approximation assumes $k_\mu \ll k_{\mu0}$ so that we can replace $u_0(\bm{k}_\mu+\bm{k}_{\mu0},\br) \approx  u_0(\bm{k}_{\mu0},\br)$.  In this case
\be \label{eqn:ema}
\langle \bm{r} \ket{\psi} \approx \sum_\mu \alpha^\mu F_\mu(\br) u_{\mu}(\br) e^{i \bm{k}_{\mu0}\cdot \br},
\ee
where $F_\mu$ is the envelope function.  Far away from the impurity  $U(\br)$ reduces to the Coulomb potential and $F_\mu$ satisfies the effective mass equation
\be \label{eqn:emt_sv}
 E\, F_\mu(\br) = \bigg[ - \frac{\hbar^2 \del_\mu^2}{2 m_{\lvert \lvert}} -  \frac{\hbar^2 \del_{\mu\perp}^2}{2 m_{\perp}} - \frac{e^2 }{4 \pi \epsilon_0 \, \epsilon\, r} \bigg] F_\mu(\br) ,
\ee 
where $\del_{z \perp}^2=\del_x^2+\del_y^2$ (similarly for other $\mu$), $m_{\lvert \lvert(\perp)} \approx 0.916 (0.191)\, m_0$ is the effective mass in the direction parallel(perpendicular) to $\hat{\mu}$, $m_0$ is the electron mass, $\epsilon_0$ is the dielectric constant, $e$ is the electron charge, and $\epsilon \approx 12$ is the static dielectric constant of Si.

Equation (\ref{eqn:emt_sv}) suggests that the ground state of the donor is six-fold degenerate; however, this degeneracy is broken by intervalley coupling induced by the donor atom.  These effects can be self-consistently included in the EMT through the multi-valley equation\cite{Ning71,Shindo76}
\begin{align} \label{eqn:emt_mv}
E \, F_\mu &=  \hat{\bm{T}}_\mu  F_\mu + \sum_{\nu }  u_\mu^* u_\nu   e^{i (\bm{k}_{\nu0}-\bm{k}_{\mu0}) \cdot \br} U(\br) F_\nu, \\
U(\br)&= -\frac{e^2 }{4 \pi \epsilon_0 \, \epsilon\, r} + U_{cc}(\br)
\end{align}
where $\hat{\bm{T}}_\mu$ is the anisotropic kinetic energy operator from Eq.\ (\ref{eqn:emt_sv}) and $U_{cc}$ is the so-called ``central cell'' potential, which takes into account deviations from the Coulombic potential in the vicinity of the donor.  Several approximation schemes have been developed to extract $U_{cc}$ based on fitting the potential by comparing  experimentally measured quantities such as energy splittings or the hyperfine coupling to the same quantities extracted from variational solutions to Eq.\ (\ref{eqn:emt_mv}).  The solutions to Eq.\ (\ref{eqn:emt_mv}) will only be reliable if the resulting $F_\mu(\bm{k})$ remain strongly localized around $\bm{k}_{\mu0}$.   Provided this constraint is satisfied, the multi-valley EMT is a powerful computatonal approach which has provided insight into the electronic structure and relaxation rates of the donors,\cite{Ning71,Shindo76,Tyuterev10} hyperfine and quadrupolar interactions of the donor nucleus, \cite{Pica14a,Mortemousque15} static Stark effects, \cite{Smit04,Friesen05} and exchange coupling between donors.\cite{Pica14b,Gamble14}


Although Eq.\ (\ref{eqn:emt_mv}) breaks the valley degeneracy, the effective Hamiltonian still commutes with the tetrahedral symmetry group $T_d$.  As a result the eigenstates of Eq.\ (\ref{eqn:emt_mv}) split into  irreducible representations of $T_d$.  The character group for the six valley sites in the reciprocal lattice splits into $\chi_{valley} = A_1 + E + T_2$.  The lowest energy state of Eq.\ (\ref{eqn:emt_sv}) is a 1$s$-like state.  The Kohn-Luttinger variational solution to $F_z$ away from the central cell is given by\cite{Kohn55}
\be \label{eqn:Fsz}
F_z \approx \frac{1}{\sqrt{\pi a b^2}} e^{  - \sqrt{ (x^2+y^2)/a^2+z^2/b^2}},
\ee
and similarly for the other $\mu$, where $a$ and $b$ are variational parameters.  The six 1$s$ states have the same symmetry as $\chi_{valley}$ and, in the orbital space, the representations are given by
\begin{align*}
\bm{\alpha}_{A_1} &= \frac{1}{\sqrt{6}}(1,1,1,1,1,1),\\
\bm{\alpha}_E &= \bigg\{\frac{1}{2}(1,1,-1,-1,0,0),\frac{1}{\sqrt{12}}(1,1,1,1,-2,-2) \bigg\},\\
\bm{\alpha}_{T_2} &= \bigg\{ \frac{1}{\sqrt{2}}(1,-1,0,0,0,0),  \frac{1}{\sqrt{2}}(0,0,1,-1,0,0), \\
& \frac{1}{\sqrt{2}}(0,0,0,0,1,-1) \bigg\}
\end{align*}
where  the six entries correspond to the $(+X,-X,+Y,-Y,+Z,-Z)$ valleys, respectively, and $\langle{\bm{r}}\ket{\psi_n}= \sum_\mu \alpha_{n}^{\mu} F_{n,\mu} u_\mu e^{i k_{\mu0} \cdot \br }$.  These irreducible representations are also important for the $2s$ states, which have a large overlap with the donor nucleus; however,  for states such as  2$p$, which vanish at the donor site, the valleys remain  decoupled and, to a good approximation, $F_{n,\mu}$ is described by Eq.\ (\ref{eqn:emt_sv}).

\section{Far-IR Spin-Valley-Orbital Control}
\label{sec:3}

In this section we show how to achieve spin-selective excitation of the donor ground state by exciting the system with far-IR light through the spin-orbit split $1sT_2$ states.  We calculate the dipole moment for this transition and estimate the achievable electron spin polarization.  This approach should be most effective for Bi donors due to the large nuclear mass and, thus, spin-orbit splitting induced by Bi.

\begin{table}[tb]
\begin{tabular}{l  c c c c} 
\hline
\hline
& P & As & Sb & Bi \\
\hline
$E_{1sT_2;\Gamma_7}-E_{1sA_1}$ (meV)\footnote{Exp., Ref.\ \onlinecite{Pajot10}.}  &11.7 &21.1 &$9.7$& $38.1$ \\
$E_{1sT_2;\Gamma_8}-E_{1sA_1}$ (meV)$^a$&- &- &10.0 & 39.1 \\

$E_{2p_0}-E_{1sA_1}$ (meV)$^a$& 34.1&42.3 &31.2 & 59.5 \\

$E_{2p_\pm}-E_{2p_0}$ (meV)$^a$&5.1 &5.1&5.1 &5.1 \\

$\lambda$  (meV)   \\
Theory &$0.02$\footnote{Theory, Ref.\ \onlinecite{Grimmeiss82}.}& $0.09^b$& $0.34^b$&$1.03^b$\\
Exp. &-&- & 0.29$^a$& 1.0$^a$ \\

$\bar{\mu}_{1sA_1,1sT_2}$ (D) \\
Theory& 0.02\footnote{Theory, this work.}  &  0.04$^c$    &0.02$^c$  &  0.05$^c$ \\
Exp.& - &- & -&$\sim 1$\footnote{Exp., Ref. \onlinecite{Krag70}, see Ref.\ \onlinecite{Pajot10}, pg.\ 180 for reproduction of data.}\\

$\bar{\mu}_{1s,2p_0}$ (D) \\
Theory & 31\footnote{Theory, Ref.\ \onlinecite{Clauws88}.} &25$^e$ &34$^e$ &15$^e$ \\

Exp. & 13\footnote{Exp., Ref.\ \onlinecite{Greenland10}.}& 10-30\footnote{Exp., Ref. \onlinecite{Jagannath81}.} &- &- \\

$\bar{\mu}_{1s,2p_{\pm}}$  (D) 
\\Theory &59$^e$ &45$^e$  &65$^e$ &28$^e$ \\
Exp. &49\footnote{Exp., Ref.\ \onlinecite{Clauws88}.}&33$^h$& - & - \\
\hline
\hline
\end{tabular}  
\caption{Table of  parameters relevant for far-IR spin-valley-orbital control.  Here $\bar{\mu}_{ab}^2 ={ \frac{1}{3} \sum_{i=x,y,z} |\bra{b}i \ket{a}|^2}$ is the average dipole moment given in Debye. \label{table:farIR} }
\end{table}

\begin{figure*}[htbp]
\begin{center}
\includegraphics[width=.75 \textwidth]{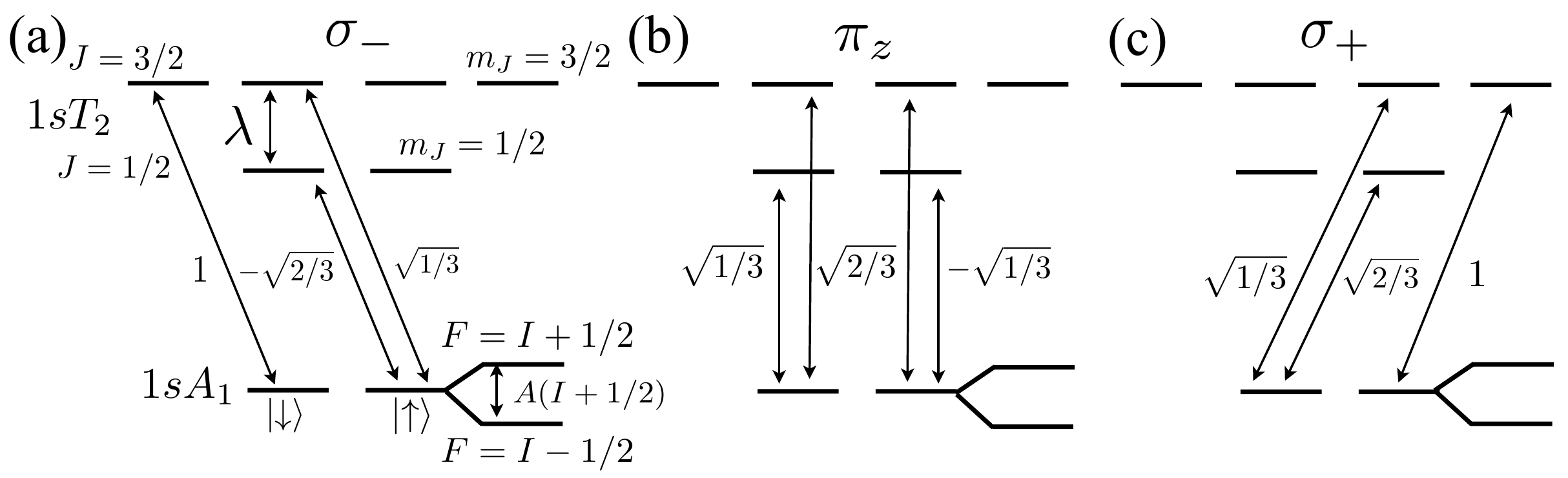}
\caption{Selection rules and Clebsch-Gordon coefficients for (a) left-hand-circularly polarized light $\sigma_-$, linearly polarized light along the $z$-axis $\pi_z$, and (c) right-hand-circularly polarized light $\sigma_+$.  We also show the two hyperfine manifolds, however, this energy splitting is typically smaller than the $1sT_2$ linewidth so is not resolvable in these transitions.}
\label{fig:selrules1}
\end{center}
\end{figure*}

The spin-orbit splitting of the excited states of the donor is negligible due to the strong dielectric screening and weak spin-orbit coupling in silicon.  However, in the vicinity of the donor nucleus, the $1s$ states  have a strong interaction with the donor nucleus, which can enhance the spin-orbit splitting.  In particular,
\be
H_{so} = \frac{\hbar^2}{2 m_0^2 c^2} \nabla U \times \bm{p} \cdot \bm{s} \approx \frac{\hbar^2}{2 m_0^2 c^2} \frac{1}{r} \frac{d U_{cc}}{dr} \bm{L} \cdot \bm{s}
\ee
where $\bm{p}$, $\bm{s}$ and $\bm{L}$ are  the electron momentum,  spin, and angular momentum, respectively.  The spin-orbit coupling matrix elements in the $1sT_2$ manifold are given by \cite{Castner67}
\begin{align}
\lambda &=\frac{3\hbar^2}{2 m_0^2 c^2} \bra{T_{x}} \frac{1}{r} \frac{dU_{cc}}{dr} L_z \ket{T_{y}} \\
& \approx \frac{3\hbar^2}{2 m_0^2 c^2} \int d^3r F_y^*F_x u_y^* u_x  \frac{1}{r} \frac{dU_{cc}}{dr} k_0y \sin k_0 y \cos k_0 x \nonumber
\end{align}
The $1sT_2$ states have a spin-one representation with respect to $\hat{\bm{L}}$ and split into  effective spin-1/2 and spin-3/2 manifold separated by $\lambda$.  The eigenstates $\ket{J,m_J}$ are characterized by the usual total angular momentum $J$ and and $z$-angular momentum $m_J$   such that\cite{Gamma78}
\begin{align} \label{eq:so_eig} 
J=1/2&:\ket{1/2,1/2}= \sqrt{2/3} \ket{T_+ \downarrow} - \sqrt{1/3}\ket{T_0 \uparrow}, \\ 
J=3/2&: \ket{3/2,3/2}= \ket{T_+ \uparrow}, \\ 
&\ket{3/2,1/2}=\sqrt{1/3} \ket{T_+ \downarrow} + \sqrt{2/3}\ket{T_0 \uparrow}, 
\end{align}
Here $\ket{T_m}$ are the projections of $\hat{L}_z$ eigenstates into the $1sT_2$ space.

Taking $F_\mu$ to be of the form of Eq.\ (\ref{eqn:Fsz}) with $a$ and $b$ given by the Kohn-Luttinger values\cite{Kohn55,Koiller01} gives an estimate that the spin orbit interaction should be reduced from the atomic value by a factor of $10^{-3}-10^{-4}$ [\onlinecite{Castner67}].  This rough estimate is consistent with the calculated values (P, As, Sb, Bi) and the measured values  (Sb, Bi) shown in Table\ \ref{table:farIR}.

\subsection{Parity-Forbidden Transitions}
Taking advantage of these spin-orbit split states for quantum control requires optical addressing of these transitions.  Unfortunately, the $1sA_1 \to 1sT_2$ transition is neither dipole nor Raman allowed within the single valley EMT.  This is because the $1s$ states are parity eigenstates within the EMT with eigenvalue $+1$, while dipole transitions should change the parity when it is a good quantum number.  At the same time, the absence of Raman transitions follows because the $1sT_2$ states are antisymmetric combinations of opposing valley states, while the $1sA_1$ states are symmetric combinations; thus a very large wavevector $\sim k_0$ is required to induce a transition between these states.

Although parity is preserved within the single valley EMT, parity is not a good quantum number within the tetrahedral group, thus, within the multi-valley EMT these transitions become dipole allowed.  The dipole matrix element is given by
\begin{align} \nonumber
\mu=\bra{T_0} z\lvert 1sA_1 \rangle &= \frac{i}{\sqrt{3}}\int d^3r ( F_{T_2,z}^{*}F_{A_1,z} z \sin 2 k_0 z \\  \label{eqn:mu} 
&+ 4  F_{T_2,z}^{*}F_{A_1,x} z \sin k_0 z\cos k_0x),
\end{align}
where we have approximated the product of Bloch functions $u_\nu^* u_{\mu} =1$ for all $\nu$ and $\mu$.
Similarly $ \bra{T_m} x\pm i y  \ket{1sA_1} /\sqrt{2} =\mu\, \delta_{m\pm} $.  The polarization selection rules for the $1sA_1$ to $1sT_2$ triplet are analogous to the case for an $s$ to $p$ transition in a spherically symmetric atom:  light polarized along the $z$ axis ($\pi_z$-light) will excite the state $T_0$, while circularly polarized light $\hat{x} \pm i \hat{y}$ ($\sigma_\pm$-light) excites the states $T_\pm$, respectively.   We show the full selection rules, including the Clebsch-Gordon coefficients in Fig.\ \ref{fig:selrules1}.

To calculate the dipole matrix elements we use the recently developed variational solutions for the $1s$ states for each donor from Ref.\ \onlinecite{Pica14a}.\cite{VarFoot} These variational solutions, together with the approximate form for $U_{cc}$, give the correct energies of the six $1s$ states for the four group-V donors and give good agreement with the  measured hyperfine coupling of the $1sA_1$ ground state.\cite{Pica14a}  The results are shown in Table \ref{table:farIR} for $\bar{\mu}_{1sA_1,1sT_2}$ defined by $\bar{\mu}_{ab}^2 ={ \frac{e}{3} \sum_{i=x,y,z} |\bra{b}i \ket{a}|^2}$.  This quantity is related to the oscillator strength  for the $a \to b$ transition 
\be \label{eqn:fab}
f_{ab} = \frac{2 m^*}{e^2 \hbar^2} (E_b - E_a) \bar{\mu}_{ab}^2, 
\ee
where ${m^*=3\, (1/m_{||} +2/m_\perp)^{-1}}$ is the average effective mass for the Si conduction band.  
In Table \ref{table:farIR} we tabulate  $\bar{\mu}_{ab}$ using measured oscillator strengths and Eq.~(\ref{eqn:fab}).  The oscillator strengths were taken from previously reported values based on absorption measurements in  doped samples.\cite{Pajot10}   
We also show the theoretical and experimental values for the $1s$ to $2p_{0,\pm}$ transitions, which are about three orders of magnitude larger.   The forbidden $1sA_1$ to $1sT_2$ transition has only been directly observed in absorption measurements on Bi doped samples,\cite{Krag70,Pajot10} which is consistent with the expectation that Bi has the largest overlap of the $1s$ states with the nucleus. (Note that this in contrast to the deep chalcogen donors in silicon, where the $1sA_1$ to $1sT_2$ transitions are more readily observable.\cite{Steger09})  However, the measured dipole moment for Bi donors is 20 times larger than what we calculate from the variational wavefunctions.  It is possible that Umklapp processes, neglected in the EMT, strongly contribute to this transition or that the variational wavefunctions are inaccurate.       
With all this taken into account, we conclude that, due to the combination of a large spin-orbit splitting and dipole element, Bi and Sb are the most promising donors for the purposes of direct optical spin manipulations through the $1sT_2$ state.

Finally, we end this section by remarking that for all the donors  it would be possible to resonantly enhance this transition using strain and electric fields via the techniques described in Sec.\ \ref{sec:4b}.  This would allow resonant manipulation via a Raman transition from $1sA_1$ to a hybridized state of $2p_0$ and $2s$ and then to the $1sT_2$ states.  As the method is similar to what we describe below, we leave the detailed analysis to Sec.\ \ref{sec:4b}.


\subsection{Electron-Nuclear Spin Polarization}
\begin{figure}[tb]
\begin{center}
\includegraphics[width=.49 \textwidth]{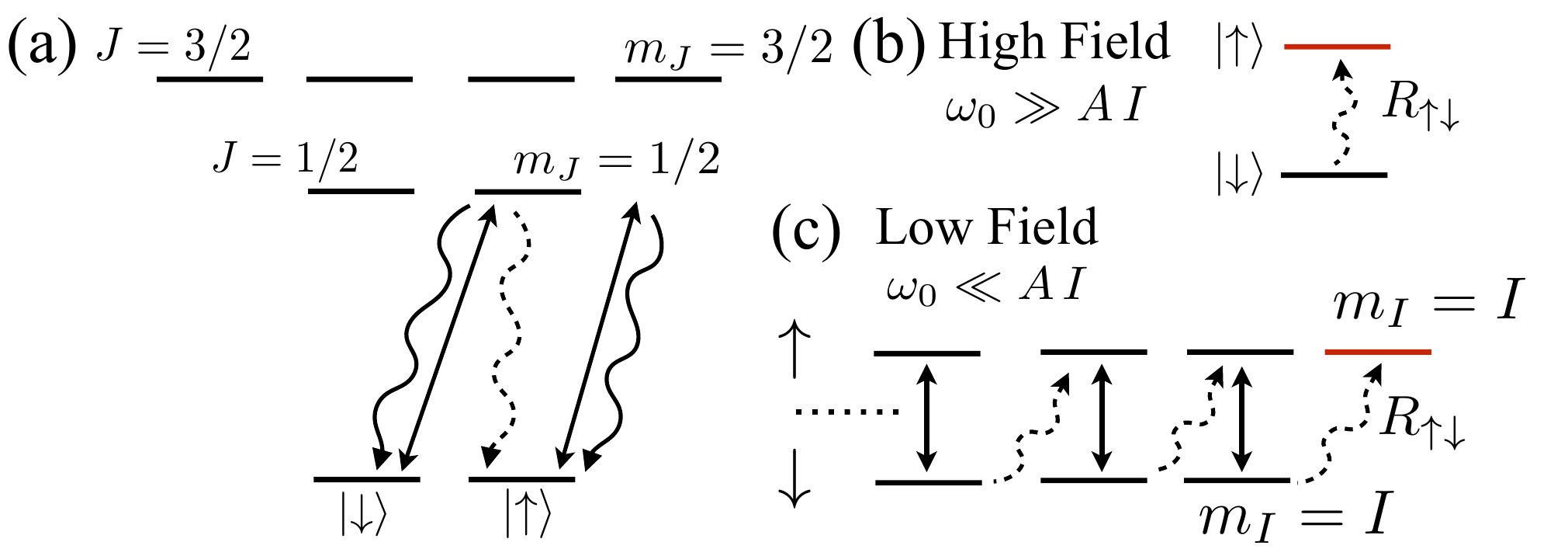}
\caption{(a) Electron spin polarization due to optical pumping with $\sigma_+$ light.  The spin-flipping optical pumping term (dashed line) requires the spin-orbit coupling to be larger than the $1sT_2$ linewidth.  (b) In the high-field regime the electron and nuclear spin are decoupled.  The electron spin is polarized into the up state (red) at the rate $R_{\uparrow\downarrow}$ [see Eq.\ (\ref{eqn:Rud})]. (c)  In the low field regime the electron and nuclear spins are admixed by the hyperfine coupling (vertical arrows).  This competes with the optical pumping to drive the system into the fully polarized electron-nuclear spin state (red).}
\label{fig:2}
\end{center}
\end{figure}

 The selection rules  shown in Fig.\ \ref{fig:selrules1} allow optical pumping of any donor into  an electron spin-polarized state by driving the system with circularly polarized light.  This process is illustrated in Fig.\ \ref{fig:2}(a--b) for $x+i y$ polarized light.\cite{QuantFoot}
In this section, we analyze this process in more detail and find the pumping rates and total electron-nuclear spin polarization as a function of the spin-orbit coupling.  We consider the high and low magnetic field regimes, which are defined by the condition that the electron Zeeman energy is much larger or smaller than the hyperfine splitting, respectively.

In the high field regime, the electron and nuclear spins decouple and we can describe the electron spin dynamics independently of the nuclear spins.  We take the control field  on resonance with the $J=1/2$ manifold.  Assuming the electron initially starts in the down state, for Rabi frequencies $\Omega = \mu E/\hbar \ll \gamma$ ($E$ is the electric field amplitude) and short times, we can make the approximation 
\begin{align}
\ket{\psi} &\approx \ket{\downarrow} + c_{-} \ket{1/2,1/2} +c_+ \ket{3/2,1/2},\\
\dot{c}_+ &= -(\gamma+ i \lambda/\hbar) c_+ + i  \Omega/\sqrt{3}, \\
\dot{c}_- &= -\gamma c_- + i \sqrt{2/3}\, \Omega,
\end{align}
where the eigenstates $\ket{J,m}$ are given in Eq.\ \ref{eq:so_eig} and $c_\pm \sim \Omega/\gamma \ll 1$.  Then the quasi-steady state is given by
\be
\ket{\psi} \approx \ket{\downarrow}+\frac{i \Omega}{\gamma} \bigg[\frac{\hbar \gamma+ i \lambda/3}{\hbar\gamma+i \lambda} \ket{T_+ \downarrow} - \frac{\sqrt{2}}{3} \frac{i \lambda}{\hbar \gamma+i \lambda} \ket{T_0 \uparrow} \bigg]
\ee
The optical pumping rate out of the state $\ket{\downarrow}$ into the state $\ket{\uparrow}$ is
\be \label{eqn:Rud}
R_{\uparrow\downarrow}=2 \gamma \abs{\langle{T_0 \uparrow} \ket{\psi } }^2  \approx \frac{4}{9}\frac{\Omega^2}{\gamma} \frac{\lambda^2}{(\hbar\gamma)^2+\lambda^2}.
\ee
Therefore the minimal requirement to have rapid optical pumping is that $\lambda \gtrsim \hbar \gamma \approx(0.01-0.1)$~meV, which is satisfied for all the donors as shown in Table \ref{table:farIR}.  

If we allow for electron spin relaxation at rate $1/T_1 \ll R_{\uparrow\downarrow}$, then this optical pumping process leads to an electron spin polarization given by 
\be
{p} = \frac{1+R_{\uparrow\downarrow}T_1}{2+R_{\uparrow\downarrow}T_1} \approx 1 -(R_{\uparrow\downarrow}T_1)^{-1}.
\ee
For Bi with resonant, far-IR light at an intensity of $10~$W/cm$^{-2}$, $\mu=1$~D,\cite{DipFoot} and $\gamma/2\pi=1.5~$GHz, we find $\Omega/2\pi = 200$~MHz and $R_{\uparrow\downarrow}/2\pi = 80$~MHz.   The $T_1$ depolarization time varies from $\approx 1$~s in nanodevices at large magnetic fields to $\gtrsim 1000$~s in bulk samples, \cite{Feher59,Wilson61,Morello10,Wolfowicz13,Steger12}, both much slower than the polarization rate.  Consequently, this optical pumping process would lead to rapid polarization of the electron spin $p>0.999$ in $\sim 100$~ns.  For P, As, and Sb, we suspect similar dynamics to Bi because our estimate for the  dipole moment is the same order of magnitude (see Table \ref{table:farIR}); however, unlike Bi, these transitions have never been directly observed in these donors.


In the low magnetic field regime, we also need to take into account hyperfine coupling.  The Hamiltonian for the electron-nuclear spin system is of the form 
\be
H_{hf}=\hbar \omega_0 s_z- \hbar\omega_n I_z + \hbar A\, \bm{s}\cdot \bm{I},
\ee
where $\bm{s}(\bm{I})$ is the donor electron(nuclear) spin operator, $I=1/2,3/2,5/2,$ and 9/2 for P, As, Sb, and Bi, respectively, $\omega_{0(n)}$ is the Zeeman energy of the electron(nuclear) spin, and $A=(117.53,198.35,186.80,1475.4)~$MHz are the hyperfine coupling constants for (P,As,Sb,Bi).   
The two fully polarized states $\ket{\uparrow, I},\ket{\downarrow,-I}$ are eigenstates for any magnetic field.

Taking the same configuration for the optical driving field as the high field regime we can assume  $R_{\uparrow\downarrow} \ll A$.  In this case, the nuclear spin will rapidly polarize with the electron spin because  the only pure electron spin up state is $\ket{\uparrow I}$.  The other $m_I$ states are admixed with the spin-down state through the hyperfine coupling and are not steady states [see Fig.\ \ref{fig:2}(b)].  In this limit, the occupation probability $p_m$ of the  states $\ket{\uparrow,m}$ is perturbatively suppressed in $1/R_{\uparrow\downarrow}T_1$, i.e., $p_{m}\sim p_{m-1}  R_{\uparrow\downarrow} T_1$.  Therefore, the electron-nuclear spin polarization $p_I$ is similar to the case without the nuclear spins, and scales as
\be
{p_I} \approx \frac{1+R_{\uparrow\downarrow}T_1}{2+R_{\uparrow\downarrow}T_1} \approx 1 -(R_{\uparrow\downarrow}T_1)^{-1}.
\ee
Similar to the high field regime, this allows for rapid polarization of the donor electron-nuclear spin system $p_I>0.999$ in $\gtrsim 100$~ns.

We end this section by noting that strain will shift the relative energies of the six valley states, which will admix the states $\ket{3/2,\pm1/2}$ with $\ket{1/2,\pm1/2}$.  When this strain coupling is much larger than $\lambda$, then the polarization process will no longer be effective because the eigenstates become pure spin states.  Using the values for the strain parameters in Si from Ref.\ \onlinecite{Yu10}, we find that for P and As the strain coupling is equal to $\lambda$ for strains around $10^{-5}$, while for Bi, this occurs around $10^{-4}$.    Strains as large as $\sim10^{-3}$ are common in Si nanodevices,\cite{Thorbeck14} indicating that this polarization process is most applicable to Bi in nanodevices, but will also be achievable in low strain environments, such as bulk samples, for the other donors.

\section{Near-IR Spin-Valley-Orbital Control}
\label{sec:4}
In this section we consider near-IR control of the donors using two-photon transitions through the spin-3/2 donor bound exciton states D$^0$X (see Fig.\ \ref{fig:1}).  We calculate the  dipole matrix elements for transitions to even-parity states and consider schemes using external electric fields and strain to couple to odd-parity states of the donor.
The D$^0$X state forms when exciton binds to the neutral donor state D$^0$.  Despite the indirect band-gap of Si, this transition is optically active because the point-like nature of the donor enables momentum conservation during photon absorption or emission. \cite{Leuenberger09}

\begin{table}[b]
\begin{tabular}{l  c c c c} 
\hline
\hline
& P & As & Sb & Bi \\
\hline
$E_{D^0X}-E_{1sA_1}$ (eV)\footnote{Exp., Ref.\ \onlinecite{Davies89}.}  &1.150 &1.149 &1.150 & 1.147 \\

$\tau$ ($\mu$s)$^a$ & 0.272 & 0.183 & - & 0.0086 \\

$\bar{\mu}_{1sA_1,D^0X}$ (D)\footnote{Exp., Ref.\ \onlinecite{Dean67b}. }& 0.033 & 0.039 & 0.033 & 0.058 \\

$\tau_r$ ($\mu$s)\footnote{Estimated from $\bar{\mu}_{1sA_1,D^0X}$.} & 1100 & 750 & 1100 & 350 \\

$\beta_2/\beta_1$  \\
Exp. & 0.24\footnote{Estimated from the ratio of photoluminescence intensity for no-phonon (P, As) or transverse-optical phonon sideband (Sb) transitions ending in $1s$ versus $2s$ reported in Ref.\ \onlinecite{Dean67a}} & 0.14$^d$ & 0.32$^d$ & - \\

Theory	& 0.06\footnote{Theory, this work (neglects central cell corrections).}  & - & - & - \\
$\beta_3/\beta_1$\\
Theory  & 0.03$^e$ & - & - & - \\
 $\beta_4/\beta_1$ \\
 Theory &	 0.02$^e$ & - & - & - \\

%

\hline
\hline
\end{tabular}  
\caption{Table of physical parameters relevant for near-IR spin-valley-orbital control.  As the free exciton  recombination energy is around $1.1545$ eV,\cite{Dean67a} we can see that the D$^0$X state has a binding energy around 5~meV (note the Si conduction band gap ${E_g \approx 1.17}$~eV), $\tau$ is the $D^0X$ state lifetime limited by Auger scattering, and $\tau_r$ is the radiative lifetime calculated from $\bar{\mu}_{1sA_1,D^0X}$.  \label{table:nearIR} }
\end{table}

In Table \ref{table:nearIR} we compile some of the relevant parameters for near-IR control of the donors.  It is important to note that, due to Auger recombination processes, these states are far from radiatively broadened and the ratio of their natural linewidth to the radiative linewidth is $\tau_r/\tau \approx 4000$ for P and As and $40\, 000$ for Bi.  As a result, creating an efficient optical interface to the donor spin states with these states is  challenging.  They have  proven to be a powerful resource for electron-nuclear spin to charge conversion for spin readout, spin control, and state initialization.\cite{Yang06,Yang09,Sekiguchi10,Lo15}  Here we explore their potential use for quantum control of the spin-valley-orbital states of the donor.

To characterize these optical transitions it is important to consider how the Coulomb interactions between electrons and holes affect the D$^0$X ground state.  To treat this problem we use a Hartree-Fock approximation with the assumption the two electron ground state is predominantly a spin-singlet in a single orbital state with  $A_1$ symmetry.\cite{Chang82,Leuenberger09}  This is a good approximation because of the large orbital splittings of the donor.  For valley electron and hole envelope functions $F_e^\nu$ and $F_h$, respectively, the 
Hartree potentials take the form \cite{Chang82,Leuenberger09}
\begin{align}
U_{eH}&=U(r)- eV_e(r)-eV_h(r),\\
U_{hH}&=\frac{e^2}{4 \pi \epsilon_0 \epsilon r} + 2 e V_e(r),\\
V_e(r)&=-\frac{e}{4 \pi \epsilon_0 \epsilon} \frac{1}{6}\sum_\nu\int d^3 r'  \frac{ \abs{F^\nu_e(r')}^2}{\abs{\bm{r}-\bm{r}'}}, \\
V_h(r)&=\frac{e}{4 \pi \epsilon_0 \epsilon} \int d^3 r' \frac{ \abs{F_h(r')}^2}{\abs{\bm{r}-\bm{r}'}}. 
\end{align}
Within the single-valley EMT, this gives rise to the Hartree-Fock (HF) equations for the two electrons and the hole
\begin{align} \label{eqn:hfe}
\varepsilon_{e} F_{e\nu}^{H}&=\big( \hat{\bm{T}}_\nu + U_{eH} \big) F_{e\nu}^{H} , \\
  \varepsilon_h F_h^{H}&=\Big( -\frac{\hbar^2 \nabla^2}{2 m_h} + U_{hH} \Big) F_h^{H}, \label{eqn:hfh}
\end{align}
where the index $\nu$ refers to the valley state and $m_h=0.49(0.16) m_0$ is the heavy(light)  hole mass.

\begin{figure*}[htbp]
\begin{center}
\includegraphics[width=.75 \textwidth]{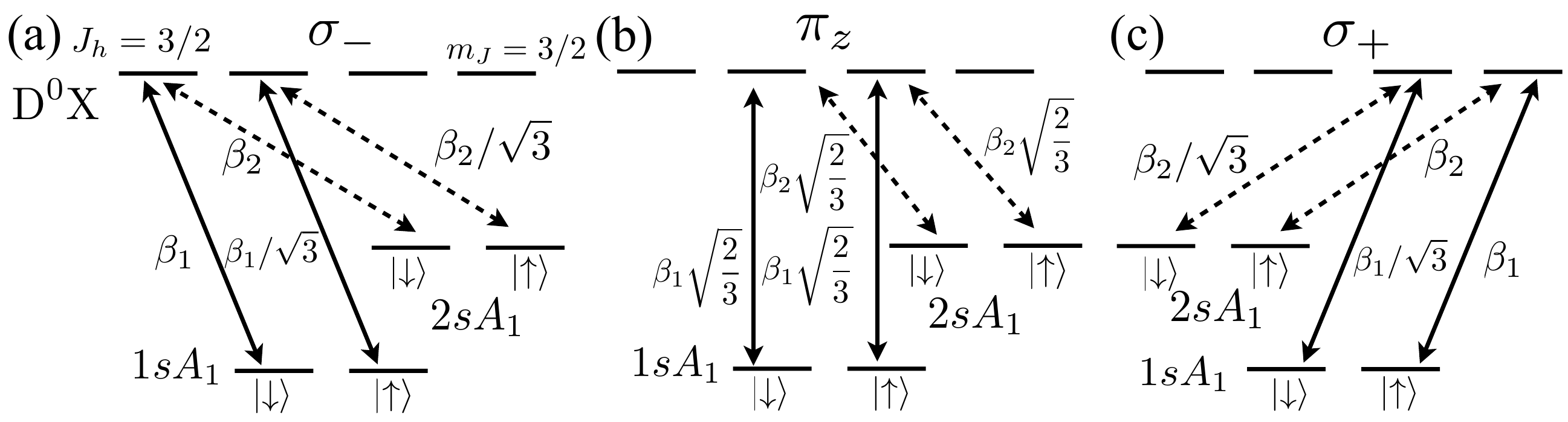}
\caption{Selection rules, Clebsch-Gordon coefficients, and overlap factors for exciting from the $1sA_1$ and $2sA_1$ states to the D$^0$X state for (a) left-hand-circularly polarized light $\sigma_-$, linearly polarized light along the $z$-axis $\pi_z$, and (c) right-hand-circularly polarized light $\sigma_+$. }
\label{fig:selrules}
\end{center}
\end{figure*}

To solve the HF equations we use a pair of variational solutions for the electrons and the holes.  
For the electrons we take a Kohn-Luttinger form for the variational wavefunction shown in Eq.~ (\ref{eqn:Fsz}), while for the hole we take a hydrogenic wavefunction of the form  \cite{Leuenberger09}
\be
F_h^{H}= \frac{1}{2 c^{5/2}} \frac{r}{\sqrt{3\pi}}e^{-r/2c},
\ee
with $c$ a variational parameter. $F_h^{H}$ vanishes at the origin, where the donor potential is repulsive, and is spherically symmetric, which is consistent with the symmetry of the valence band and the symmetric form of $U_{hH}$.   To find the variational parameters we fix $c$ and minimize the expectation value of Eq.\ (\ref{eqn:hfe}) with respect to $a$ and $b$, and similarly for the hole, we fix $a$ and $b$ and minimize  the expectation value of Eq.\ (\ref{eqn:hfh}) with respect to $c$.   Self-consistent variational solutions occur when these two minimization procedures produce the same values for $a,$ $b$, and $c$.  For the heavy hole we find $a=2.79$~nm, $b=1.60$~nm, and $c=1.34$~nm, while for the light hole we find $a=2.74$~nm, $b=1.57$~nm, and $c=1.99$~nm.  This is to be compared to the case for the $D^0$ state where the Kohn-Luttinger values are $a=2.51$~nm and $b=1.44$~nm.\cite{Koiller01}


\subsection{Two-Photon Transitions to Even-Parity States}
We now explore the potential for spin-selective Raman transitions from the $1sA_1$ ground state to the even parity excited orbital states of the donor such as the $2sA_1$ and $3sA_1$ states.   The two electrons in the $D^0X$ ground state are expected to primarily pair in a spin-singlet in the $1sA_1$ orbital, as a result the $nsA_1$ orbitals will become admixed with the ground state through the Coulomb interactions between the electrons and holes.  Such two-photon transitions have been used to explain some of the satellite emission lines observed in photoluminescence spectra of the donor bound excitons.\cite{Dean67a,Davies89}  In Table \ref{table:nearIR} we list the ratio of the dipole moments for the no-phonon transitions $D^0X\to1sA_1$ vs the $D^0X \to 2sA_1$ states based on these early measurements.\cite{Dean67a}   

We can also calculate this ratio using the Hartree-Fock theory.  Our variational solution has the same symmetry as the $ns$ states of the singly occupied donor, which allows the expansion
\begin{align}
\ket{\psi_{e\nu}^{H}}&= \sum_{n} \beta_n \ket{\psi_{ns}^\nu},\\
\beta_n &=\langle \psi_{ns}^\nu \ket{\psi_{e\nu}^H}=  \int dr F_{ns}^{\nu*}(r) F_{e\nu}^H (r).
\end{align}
We approximate the $ns$ states by an orthogonal set of hydrogen-like wavefunctions \cite{Faulkner69}
\begin{align}
F_{ns}^\nu& =\frac{1}{\sqrt{\pi\, n^5 a^2 b}} L_{n-1}^{(1)}(2 \rho/n)  e^{- \rho/n}, \\
\rho& = \sqrt{(x^2+y^2)/a^2+z^2/b^2},
\end{align}
where $L_{n-1}^{(1)}$ are the generalized Laguerre polynomials for the $ns$ states and we take the same values for $a$ and $b$ as the $1s$ state to ensure orthogonality.  The ratio of the dipole matrix elements is then given by
\be
\frac{\mu_{n'sA_1,D^0X}}{\mu_{nsA_1,D^0X}} = \frac{\beta_{n'}}{\beta_n}.
\ee
In Table \ref{table:nearIR} we tabulate these ratios for the first few $1s$ states.  In general, we find $\beta_n \sim 1/n$ as $n$ increases.  We show the full selection rules, including the Clebsch-Gordon coefficients, for these transitions in Fig.\ \ref{fig:selrules}

Additional two-electron transitions have also been observed to the $1sE$ and $1sT_2$ states.\cite{Davies89}   However, these states become admixed with the $D^0X$ ground state through valley-orbit interactions, which we have not included in the Hartree-Fock analysis.  

\subsection{Two-Photon Transitions to Odd-Parity States} \label{sec:4b}
In this section, we show how to achieve two-photon transitions to the $np$-states by applying either a static electric field or strain to the donor system.  This additional control is required because
the hydrogenic model is a good approximation for the donors and parity is a good quantum number.   Parity can be broken by applying a static electric field, which will mix $ns$ and $np$-like states.  We analyze this case in Sec.\ \ref{sec:4b1} below.  Alternatively, valley-orbit interactions can break the parity symmetry because the tetrahedral group does not conserve parity.  Similar to the case in Sec.\ \ref{sec:3}, there is a direct transition from the $D^0X$ state to the $2p^0$ states which are antisymmetric combinations of the same valley states.  This occurs with the relative dipole moment 
\be \label{eqn:oddPar}
\frac{\mu_{2p_0,D^0X}}{\mu_{1sA_1,D^0X}}\approx  i\int d^3r\,  u_{-z}^* u_z F_{2p_0}^{ z*} F_{ez}^{H} \sin k_0 z,
\ee
Unfortunately, this is suppressed by a factor of $\sim10^3$ compared to the $nsA_1$ transitions, due to the rapidly varying Bloch phase.  In Sec.\ \ref{sec:4b2}, we show that these transitions are resonantly enhanced for certain values of strain when the $2s$ states are degenerate with the $2p$ states to within $\sim(10-100)~\mu$eV.

\subsubsection{Static Electric Field} \label{sec:4b1}
When an electric field $\bm{E}$ is applied to the donor the single-valley EMT is modified to
\be
 E F_\mu(\br)=\Big[  \hat{\bm{T}}_\mu +U(r) + e \bm{E}\cdot \bm{r} \Big] F_\mu(\br).
\ee
By analogy with Stark effect for a hydrogen atom, we can write an ansatz for the $2p_0$ state in the $\pm k_{0z}$ valley \cite{Kohn55,Pica14a}
\be
F_{2p_0}^{\pm z} = N z (1+ q z) e^{-\sqrt{(x^2+y^2)/a^2+z^2/b^2}},
\ee
where $N$ is a normalization constant, $q$, $a$, and $b$ are variational parameters and we took $\bm{E}$ parallel to the $z$-valley axis.  For each value of $\abs{\bm{E}}$ we minimize the energy expectation value to find the variational parameters.
For finite $q$ the symmetric combination of these valley states has a direct overlap with $1sA_1$ Hartree-Fock orbital, which results in the ratio of dipole moments
\be
\frac{\mu_{2p_0z,D^0X}}{\mu_{1sA_1,D^0X}}= \frac{1}{\sqrt{3}} \int d^3r F_{2p_0}^{ z*} F_{ez}^{H} .
\ee
For small fields, we are  justified in taking the zero-field envelope function for the Hartree-Fock orbital because of the  weak dipole moments of the $D^0X$ state compared to the $2p_0$ state.  

\begin{figure}[t]
\begin{center}
\includegraphics[width=.4\textwidth]{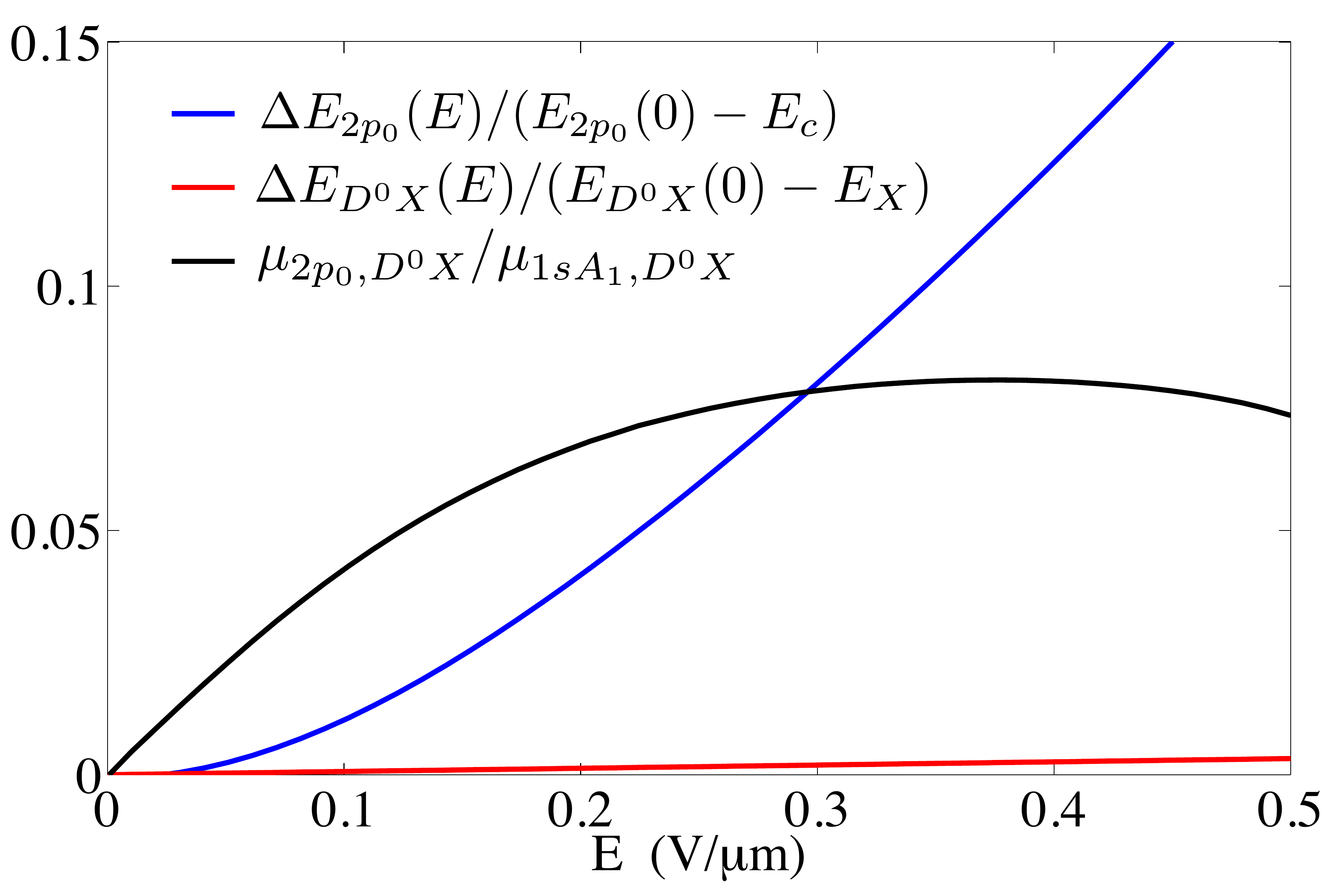}
\caption{(Black) Relative dipole moment for optical transitions from $D^0X$ ending in the odd-parity $2p_0$ state with increasing electric field, which breaks the parity symmetry.  (Blue/Red) Energy shifts of $2p_0/D^0X$ states relative to their binding energy with applied electric field.  The peak in the dipole moment occurs when both shifts are relatively small.  For small fields the dipole moment is small because parity is only weakly broken, while for large electric fields the $2p_0$ state becomes further shifted away from the donor site where it has a large overlap with the $1sA_1$ state.  }
\label{fig:4b1}
\end{center}
\end{figure}

In Fig.\ \ref{fig:4b1} we show this ratio under the application of experimentally relevant electric fields (note the the donors are ionized at electric fields around $2~V/\mu$m\cite{Pica14a}).  We also plot the relative shift in the energy of the $2p_0$ state and the $D^0X$ state with applied electric field.   For the $D^0X$ we used the Stark shift parameter $2 p_8 \approx 33~\mu$eV/$(\textrm{V}/\mu\textrm{m})^{-1}$  measured in Ref.\ \onlinecite{Lo15}.  The shift in the $D^0X$ state is just a few-percent of the binding energy, indicating the zero-field envelope is a good approximation to the wavefunction.   The shift in the $2p_0$ energy is much larger, but it is still small compared to the $2p_0$ binding energy.  From the figure we see that at the optimal value of the electric field, the ratio of the dipole moment to the odd-parity $2p_0$ state is the same order as the ratios we found for the even-parity transitions at zero-field.  Similar arguments also hold for the various $2p_\pm$ states.  Thus we can conclude that a static electric field  is a realistic approach for achieving  two-photon transitions to these odd-parity valley-orbit states.

\subsubsection{Strain Enhanced Valley-Orbit Interactions}  \label{sec:4b2}
\begin{figure}[t]
\begin{center}
\includegraphics[width=.49\textwidth]{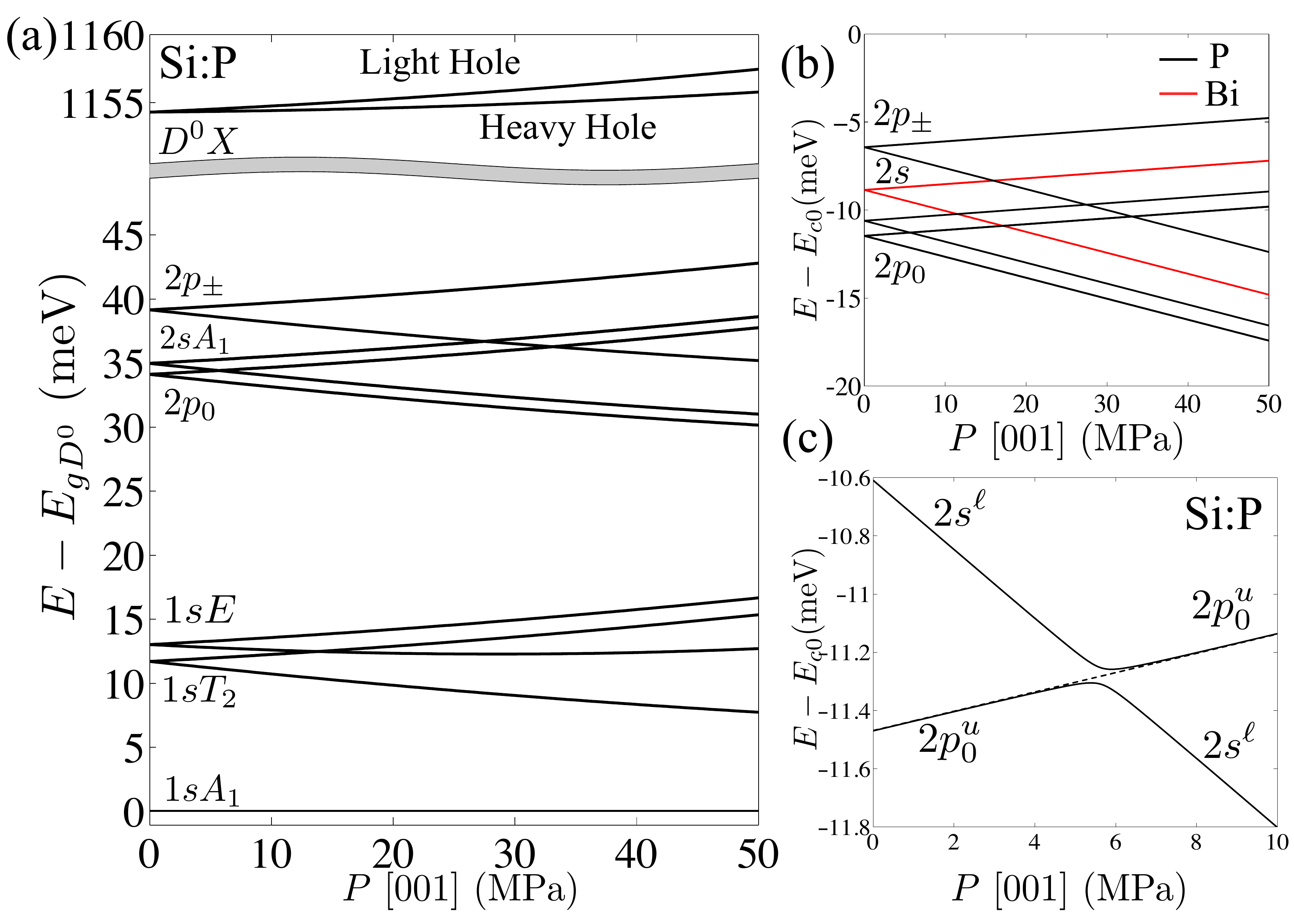}
\caption{(a) Spectrum of donor transitions for Si:P from the $D^0$ ground state to the excited $D^0$ states and the $D^0X$ state with increasing compressive force along the [001] axis.  Note for P the spin-orbit splitting of the $1sT_2$ and $2p$ states is negligible compared to the linewidth of these states.  (b) Energies of  $2p$ and $2s$ states with energy measured relative to the conduction band at zero detuning.  Resonances appear at specific strain values.  Also shown are the states for Si:Bi where resonances appear between $2s$ and $2p_0$ and $2p_\pm$ at similar pressures. (c) Avoided crossing between the symmetric combination of lower-valley $2s^\ell$ states  and the four upper-valley $2p_0^u$ states for $\Delta = 10~\mu$eV.  Two of the five states repel each other due to valley-orbit interactions, while the other three are unaffected.  }
\label{fig:4b2}
\end{center}
\end{figure}

We now show that it is possible to achieve comparable two-photon transitions to the $2p$ states without applying an additional electric field.  In Fig.\ \ref{fig:4b2} we show the shift in the energy levels of the $D^0$ and $D^0X$ states with a compressive pressure along the $[001]$ axis  for phosphorous donors using the model developed in Ref.\ \onlinecite{Lo15}.   For pressures around  5(30)~MPa we see that the lower(upper) valleys of the $2s$ states becomes resonant with the upper(lower) valleys of the $2p_{0(\pm)}$ states.   In Fig.\ \ref{fig:4b2}(b) we also show the $2sA_1$ state for bismuth donors, which has the coincidence that  these resonances occur near the same value of the pressure around (15-20)~MPa. 

Within the single-valley EMT these states remain decoupled even under the application of a small electric field because the electron is localized in opposite valleys.  However, when the valley-orbit interactions are included, these become avoided crossings because the donor potential can induce transitions between valleys.  Looking more closely at the $2s^\ell-2p_0^u$ crossing ($\ell(u)$ refers to the lower(upper) valley states under the application of strain) we can express these states as 
\begin{align} \label{eqn:upper}
\bm{\alpha}_{2s^\ell,\pm}&=\frac{1}{\sqrt{2}}(0,0,0,0,1,\pm1),\\
\bm{\alpha}_{2p_0^u,\pm x}&=\frac{1}{\sqrt{2}}(1,\pm1,0,0,0,0),\\
\bm{\alpha}_{2p_0^u,\pm y}&=\frac{1}{\sqrt{2}}(0,0,1,\pm1,0,0), \label{eqn:lower}
\end{align}
The valley-orbit interactions are given by the matrix elements
\be
\begin{split}
\Delta_{n,n'} &= \bra{ \psi_n} U \ket{\psi_n'}.
\end{split}
\ee
Because of the odd-parity of the $2p_{0}$ states and the even parity of the $2s$ states, this will only be nonzero for the states  $\alpha_{2s^\ell,+}$ and $\alpha_{2p_0^u,-x(-y)}$, in which case
\be
\Delta= 2 i \int d^3r\, u_z^* u_{x}  F_{2s}^{z*}(r) U(r) F_{2p_0}^x(r) \cos k_0 z\,  \sin k_0 x 
\ee
A simple estimate suggests this should be $\sim10^{-3}$ times the valley-orbit splitting of the $1s$ states, which implies that it can be as large as $(10-40)~\mu$eV, depending on the donor.  This is comparable to the lifetime of these states.  In Fig.\ \ref{fig:4b2}(c) we show a small region of pressure around this avoided crossing for phosphorous.  We see that the width of the avoided crossing corresponds to about $1~$MPa, indicating that it would be possible to stabilize the system at this point with suitable control of the stress.  At the avoided crossing when $E_{2s^\ell,+}=E_{2p_0^u}$ the eigenstates are given by
\begin{align}
\Delta E=\Delta:& \, \bm{\alpha}= \frac{1}{\sqrt{6}}(1,-1,1,-1;1,1),\\
\Delta E=0:&\, \bm{\alpha}=\Big\{ \frac{1}{\sqrt{2}}(1,1,0,0;0,0), \\ 
&\,\frac{1}{\sqrt{2}}(0,0,1,1;0,0), \nonumber
\frac{1}{2}(1,-1,-1,1;0,0) \Big\},\\
\Delta E=-\Delta&: \, \bm{\alpha}=\frac{1}{\sqrt{6}}(-1,1,-1,1;1,1),
\end{align}
where $\Delta E= E- E_{2p_0^u}$ and it is should be understood that the first four entries are associated with the $2p_0^u$ valley envelopes and the last two entries with the $2s^\ell$ valley envelopes.  At this resonance,  the ratio of the dipole moments will then be given by
\be
\frac{\mu_{2p_0,D^0X}}{\mu_{1sA_1,D^0X}}= \frac{1}{3} \int d^3r F_{2s}^{ z*} F_{ez}^{H} =\frac{1}{3} \frac{\beta_2}{\beta_1},
\ee
where we took the zero-stress form for the Hartree-Fock solution, which is valid approximation for these small stresses because the ground state wavefunction is protected from deforming by the large valley-orbit splitting.  This results in an enhancement of the dipole moment by a factor of $\sim10^3$ compared to the Eq.\ (\ref{eqn:oddPar}).  

We end this section by noting that although strain is common in silicon nanodevices  (it can be as large as $\sim10^{-3}$ corresponding to $P \approx 200$~MPa\cite{Thorbeck14}), the approach described here would require additional static tuning of the strain to bring the donor states near these resonances.  Furthermore, in this discussion we have only considered the influence of axial strain,  in realistic nanodevices there is also a contribution from shear strain.  Shear strain leads to additional energy shifts and it changes the effective masses of the conduction band.\cite{Ungers07}  This latter effect will change the valley splittings, but will not qualitatively change the character of these resonances.  Therefore, the main challenge for exploiting these resonances in nanodevices will be achieving the resonant strain condition in a deterministic manner.  Alternatively, mixing between $s$ and $p$ states might be induced by other effects such as proximity to an interface.  In bulk samples, these resonances should be more readily achievable by applying a global stress to the sample, as we have assumed in this section.

\section{Conclusions}
\label{sec:5}
We have shown how to use far-IR or near-IR optical fields to achieve spin selective excitation of the group-V donors from the ground state to  excited orbital states.  In the case of far-IR light, we calculated the dipole moments for transitions which are parity-forbidden within the hydrogenic approximation for the donors, but are allowed due to valley-orbit interactions.  These transitions have only been directly observed in the case of the Bi donor.  For Bi, our calculated dipole moment disagrees is two orders of magnitude smaller than the estimate based on these absorption measurements.  This suggests that further experimental and theoretical work is required to account for such a large discrepancy.   We then showed that these transitions obey selection rules consistent with excitation from a $J=1/2$ ground state to spin-orbit split $J=3/2$ and $J=1/2$ excited states.  This enables optical pumping into electron-nuclear spin polarized states through the application of circularly polarized light.  We calculated the timescale and final polarization for this process and found it is most promising for Bi donors, due to their large spin-orbit splitting and strong valley-orbit interactions.  

For near-IR light one can make use of the $J=3/2$ donor bound exciton state D$^0$X, which has weak, dipole allowed transitions from the donor ground state to D$^0$X.  We used a Hartree-Fock variational solution to the electron-hole wavefunctions for this state.  Using this solution we estimated the relative strength of the transition from the $1sA_1$ ground state to D$^0$X versus the transitions from the $nsA_1$ excited states  to D$^0$X.  For the $1s$ versus $2s$ transitions, our calculated values agree with previous experimental measurements within a factor of five.  Resonant excitation on these two transitions would enable spin-selective, two-photon, $\Lambda$-transitions from the ground state to the $nsA_1$ excited states.   We then showed that one can achieve similar $\Lambda$-transitions to $2p$ states by breaking the parity symmetry present within the hydrogenic approximation for the donors.  We considered two approaches.  In the first approach, an applied electric field directly breaks the parity symmetry and admixes the $nsA_1$ states with $2p$ states.  The second approach uses strain to bring the $2s$ and $2p$ states into resonance.  In this case, the valley-orbit interaction is resonantly enhanced and the antisymmetric combination of $2p$ valley states becomes strongly admixed with the symmetric combination of $2s$ valley states.  This enhances the $\Lambda$-transition from the $1sA_1$ to the $2p$ states by a factor of $\sim10^3$ at zero applied electric field.

We anticipate that these results will help guide future spectroscopic investigation of the donors.  In particular, many of the effects investigated here probe the poorly understood interaction between the donor bound electrons and holes and the donor nucleus.  For quantum information applications, these results provide a path forward for achieving full optical quantum control and  state preparation of group-V donors, as well as a coherent interface between donor spin qubits and single photons.\cite{Gullans14}  

\emph{Acknowledgments --} 
We thank  G.\ Wolfowicz, J.\ J.\ L.\ Morton, K. D. Greve, and S. Lyon for helpful discussions, M.\ L.\ W.\ Thewalt and M. Friesen for several insightful comments on the manuscript, and G. Pica and B. Lovett for pointing out several corrections to our calculation of  the parity forbidden dipole matrix element in Sec.~IIIA.  
Funding is provided by NIST and the  NSF Physics Frontier at the JQI.

\bibliographystyle{../apsrev-nourl}

\bibliography{../SiVGraph}

\begin{thebibliography}{53}
\bibitem[{Gam()}]{Gamma78}
\bibinfo{note}{We refer to these as $J=1/2(3/2)$ states for convenience, these
  are more accurately referred to as $\Gamma_{7(8)}$ states due to the presence
  of the tetrahedral symmetry.}
\bibitem[{Var()}]{VarFoot}
\bibinfo{note}{These variational solutions contain five variational parameters,
  which take distinct values for the $1sA_1$ state and $1sT_2$ state. See Ref.\
  \onlinecite{Pica14a} for more details.}
\bibitem[{Qua()}]{QuantFoot}
\bibinfo{note}{Note that we have rather arbitrarily chosen a quantization axis
  along the crystallographic $z$ direction. In general, the quantization
  direction will be set by a combination of the magnetic field, polarization of
  the light, and the local strain in the crystal.}
\bibitem[{Dip()}]{DipFoot}
\bibinfo{note}{Here we use the experimentally measured value for the dipole
  moment from Table \ref{table:farIR}.}
\end{thebibliography}

\end{document}